# Plasmons in nearly touching metallic nanoparticles: singular response in the limit of touching dimers

Isabel Romero,[1] Javier Aizpurua,[1] Garnett W. Bryant,[2] F. Javier García de Abajo[1,3]

[1]*Donostia International Physics Center (DIPC), Apartado 1072, 20080 San Sebastian, Spain*
[2]*National Institute of Standards and Technology, Gaithersburg, Maryland 20899-8423, USA*
[3]*Instituto de Óptica, CSIC, Serrano 121, 28006 Madrid, Spain*

**Abstract:** The response of gold nanoparticle dimers is studied theoretically near and beyond the limit where the particles are touching. As the particles approach each other, a dominant dipole feature is observed that is pushed into the infrared due to interparticle coupling and that is associated with a large pileup of induced charge in the interparticle gap. The redshift becomes singular as the particle separation decreases. The response weakens for very small separation when the coupling across the interparticle gap becomes so strong that dipolar oscillations across the pair are inhibited. Lower-wavelength, higher-order modes show a similar separation dependence in nearly touching dimers. After touching, singular behavior is observed through the emergence of a new infrared absorption peak, also accompanied by huge charge pileup at the interparticle junction, if initial interparticle-contact is made at a single point. This new mode is distinctly different from the lowest mode of the separated dimer. When the junction is made by contact between flat surfaces, charge at the junction is neutralized and mode evolution is continuous through contact. The calculated singular response explains recent experiments on metallic nanoparticle dimers and is relevant in the design of nanoparticle-based sensors and plasmon circuits.

©2006 Optical Society of America

**References and links**

1. M. Faraday, "On the color of colloidal gold," Philos. Trans. Royal Soc. London **147**, 145-181 (1857).
2. U. Kreibig and M. Vollmer, *Optical Properties of Metal Clusters* (Springer-Verlag, Berlin, 1996).
3. S. Schultz, D. R. Smith, J. J. Mock, and D. A. Schultz, "Single-target molecule detection with nonbleaching multicolor optical immunolabels," Proc. Natl. Acad. Sci. **97**, 996-1001 (2000).
4. E. Prodan, C. Radloff, N. J. Halas, and P. Nordlander, "A hybridization model for the plasmon response of complex nanostructures," Science, **302**, 419-422 (2003).
5. J. Aizpurua, P. Hanarp, D. S. Sutherland, M. Käll, G. W. Bryant, and F. J. García de Abajo, "Optical properties of gold nanorings," Phys. Rev. Lett. **90**, 057401 (2003).
6. K. G. Thomas, S. Barazzouk, B. I. Ipe, S. T. S. Joseph, and P. V. Kamat, "Uniaxial plasmon coupling through longitudinal self-assembly of gold nanorods," J. Phys. Chem. B, **108**, 13066-13068 (2004).
7. M. El-Kouedi, and C. A. Foss, "Optical properties of gold-silver iodide nanoparticle pair structures," J. Phys. Chem. B **104**, 4031-4037 (2000).
8. T. Ung, T., L. M. Liz-Marzán, and P. Mulvaney, "Optical properties of thin films of Au@$SiO_2$ particles," J. Phys. Chem. B **105**, 3441-3452 (2001).
9. H. Tamaru, H. Kuwata, H. T. Miyazaki, and K. Miyano, "Resonant light scattering from individual Ag nanoparticles and particle pairs," Appl. Phys. Lett. **80**, 1826-1828 (2002).
10. W. Rechberger, A. Hohenau, A. Leitner, J. R. Krenn, B. Lamprecht, and F. R. Aussenegg, "Optical properties of two interacting gold nanoparticles," Opt. Commun. **220**, 137-141 (2003).
11. T. Atay, J. H. Song, and A. V. Nurmikko, "Strongly interacting plasmon nanoparticle pairs: from dipole-dipole interaction to conductively coupled regime," Nano Lett. **4**, 1627-1631 (2004).
12. L. Gunnarsson, T. Rindzevicius, J. Prikulis, B. Kasemo, M. Käll, S. Zou, and G. C. Schatz, "Confined plasmons in nanofabricated single silver particle pairs: experimental observations of strong interparticle interactions," J. Phys. Chem. B **109**, 1079-1087 (2005).

13. C. E. Talley, J. B. Jackson, C. Oubre, N. K. Grady, C. W. Hollars, S. M. Lane, T. R. Huser, P. Nordlander, and N. J. Halas, "Surface-enhanced Raman scattering from individual Au nanoparticles and nanoparticle dimer substrates," Nano Lett. **5**, 1569-1574 (2005).
14. R. Rupin, "Surface modes of two spheres," Phys. Rev. B **26**, 3440-3444 (1982).
15. M. Schmeits and L. Dambly, "Fast-electron scattering by bispherical surface-plasmon modes," Phys. Rev. B **44**, 12706-12712 (1991).
16. A. V. Vagov, A. Radchik, and G. B. Smith, "Optical response of arrays of spheres from the theory of hypercomplex variables," Phys. Rev. Lett. **73**, 1035-1038 (1994).
17. A. V. Radchik, A. V. Paley, G. B. Smith, and A. V. Vagov, "Polarization and resonant absorption in intersecting cylinders and spheres," J. Appl. Phys. **76**, 4827-4835 (1994).
18. G. B. Smith, W. E. Vargas, G. A. Niklasson, J. A. Sotelo, A. V. Paley, and A. V. Radchik, "Optical properties of a pair of spheres: comparison of different theories," Opt. Commun. **15**, 8-12 (1995).
19. J. Aizpurua, A. Rivacoba, N. Zabala, and F. J. García de Abajo, "Collective excitations in an infinite set of aligned spheres," Surf. Sci. **402-404**, 418-423 (1998).
20. F. J. García de Abajo and A. Howie, "Relativistic electron energy loss and electron-induced photon emission in inhomogeneous dielectrics," Phys. Rev. Lett., 80, 5180-5183 (1998); "Retarded field calculation of electron energy loss in inhomogeneous dielectrics," Phys. Rev. B **75**, 115418 (2002).
21. F. J. García de Abajo, "Interaction of raidation and fast electrons with clusters of dielectrics: a multiple scattering approach," Phys. Rev. Lett. **82**, 2776-2779 (1999).
22. A. Pack, M. Hietschold, and R. Wannemacher, "Failure of local mie theory: optical spectra of colloidal aggregates," Opt. Commun. **194**, 277-287 (2001).
23. H. Xu and M. Käll, "Surface-plasmon-enhanced optical forces in silver nanoaggregates," Phys.Rev. Lett. **89**, 246802 (2002).
24. A. A. Lalayan, K. S. Bagdasaryan, P. G. Petrosyan, Kh. V. Nerkararyan, and J. B. Ketterson, "Anomalous field enhancement from the superfocusing of surface plasmons at contacting silver surfaces," J. Appl. Phys. **91**, 2965-2968 (2002).
25. P. Nordlander, C. Oubre, E. Prodan, K. Li, and M. I. Stockman, "Plasmon hybridization in nanoparticle dimers," Nano Lett. **4**, 899-903 (2004).
26. J. J. Xiao, J. P. Huang, and K. W. Yu, "Optical response of strongly coupled metal nanoparticles in dimer arrays," Phys. Rev. B **71**, 045404 (2005).
27. J. Aizpurua, G. W. Bryant, L. J. Richter, F. J. García de Abajo, B. K. Kelly, and T. Mallouk, "Optical properties of coupled metallic nanorods for field-enhanced spectroscopy," Phys. Rev. B **71**, 235420 (2005).
28. P. Nordlander and E. Prodan, "Plasmon hybridization in nanoparticles near metallic surfaces," Nano Lett. **4**, 2209-2213 (2004).
29. D. A. Genov, A. K. Sarychev, V. M. Shalaev, and A. Wei, "Resonant field enhancements from metal nanoparticle arrays," Nano Lett. **4**, 153-158 (2004).
30. E. D. Palik, *Handbook of Optical Constants of Solids* (Academic Press, New York, 1985).
31. F. Claro, "Absorption spectrum of neighboring dielectric grains," Phys. Rev. B **25**, 7875-7876 (1982).
32. F. Claro, "Multipolar effects in particulate matter," Sol. St. Comm. **49**, 229-232 (1984).
33. A. O. Govorov, S. A. Studenikin, and W. R. Frank, "Low-frequency plasmons in coupled electronic microstructures," Phys. Solid State **40**, 499-502 (1998).
34. F. J. García de Abajo and J. Aizpurua, "Numerical simulation of electron energy loss near inhomogeneous dielectrics," Phys. Rev. B **56**, 15873-15884 (1997).
35. L. C. Davis, "Electrostatic edge modes of a dielectric wedge," Phys. Rev. B **14**, 5523-5525 (1976).
36. D. R. McKenzie and R. C. McPhedran, "Exact modelling of cubic lattice permittivity and conductivity," Nature **265**, 128-129 (1977).
37. C. Pecharromán and J. S. Moya, "Experimental evidence of a giant capacitance in insulator-conductor composites at the percolation threshold," Adv. Mat. **12**, 294-297 (2000).
38. C. Pecharromán, F. Esteban-Betegón, J. F. Bartolomé, S. López-Esteban, and J. S. Moya, "New percolative $BaTiO_3$-Ni composites with a high and frequency-independent dielectric constant ($\varepsilon_r \approx 80\,000$)," Adv. Mat. **13**, 1541-1544 (2001).
39. C. C. Chen and Y. C. Chou, "Electrical-conductivity fluctuations near the percolation threshold," Phys. Rev. Lett. **54**, 2529-2532 (1985).
40. R. Fuchs and F. Claro, "Multipolar response of small metallic spheres: nonlocal theory," Phys. Rev. B **35**, 3722-3727 (1987).
41. S. A. Maier, P. G. Kik, H. A. Atwater, S. Meltzer, E. Harel, B. E. Koel, and A. A. G. Requicha, "Local detection of electromagnetic energy transport below the diffraction limit in metal nanoparticle plasmon waveguides," Nature Mater. **2**, 229-232 (2003).
42. L. A. Sweatlock, S. A. Maier, H. A. Atwater, J. J. Penninkhof, and A. Polman, "Highly confined electromagnetic fields in arrays of strongly coupled Ag nanoparticles," Phys. Rev. B **71**, 235408 (2005).
43. J. A. Dionne, L. A. Sweatlock, H. A. Atwater, and A. Polman, "Plasmon slot waveguides: towards chip-scale propagation with subwavelength-scale localization," Phys. Rev. B **73**, 035407 (2006).

## 1. Introduction

Since the pioneering work of Faraday on the color of colloidal gold [1], plasmons in nanoparticles have received considerable attention [2]. An increasing number of applications are being devised, such as markers for biomolecules spotted via plasmonic response [3]. Control over the plasmon frequencies has been gained by playing with particle shapes, as recently shown for nanoshells [4], nanorings [5], and nanorods [6]. Furthermore, mixing, splitting, and frequency shifting of the plasmon modes of individual particles have been experimentally [7-13] and theoretically [9,12-27] addressed in particle dimers, giving rise to the concept of plasmon chemistry [25,28]. In particular, coupling particle dipole modes have been observed to yield a strongly red-shifted dipole-active mode for external fields oriented along the inter-particle axis and a blue shifted mode for perpendicular orientations [10-12]. We shall be interested in parallel fields in this work, which produce considerably larger shifts owing to stronger mode coupling.

For coupled nanoparticle dimers there are three regimes of particle separation that exhibit distinctly different behavior [27]. For very widely separated nanoparticles, they behave independently. The only vestige of the interparticle coupling is the oscillatory variation of the response with interparticle separation (if this is not small compared to the wavelength) that arises from interference between the two nanoparticles [27]. For particle dimensions much smaller than the wavelength, interference effects become unimportant but interparticle coupling becomes significant. As the particle separation is reduced further, the dipolar response redshifts. Initially, the far-field response weakens because the effective dipole of the coupled dimer is reduced [27]. For even smaller separations, the modes continue to redshift, there is a strong buildup of charge at the gap, near-fields in the gap are enhanced, and the far-field scattering begins to increase again [27]. In this letter, we show theoretically that there are two additional regimes of distinctly different behavior. One new regime occurs when the particles are separated but nearly touching. The second new regime appears when the nanoparticles overlap. Providing an understanding of these two regimes is important for explaining recent experiments on nearly touching and overlapping dimers [11] and for designing better plasmon-based sensors and circuits.

Dimers in the near-touching limit present a challenging situation that is accompanied by (i) dramatic enhancement of the induced electric field at the point of maximum proximity [23,29], (ii) shift of plasmon resonances towards the infrared [9,11,12], (iii) the possibility of a singular transition when the particles touch,[11] and (iv) modes appearing and disappearing near touching [11]. Several theoretical studies have focused on the interaction between neighboring separated spheres both in the non-retarded limit [14-19,26] and including full retardation effects [20,21,23-25,27]. However, particles in the nearly-touching limit deserve further consideration to understand the singular transition observed in recent experiments [11], which includes strong mode shifts towards the infrared before touching and shifts away from the infrared as the particles overlap, with a discontinuous change in the evolution of the response when the particles are just touching. It is essential that barely separated dimers and just overlapping pairs be treated on a common footing, as done here, to have a consistent description of the mode transition through touching.

In this theoretical work, we discuss in detail how plasmon modes strongly shift toward the infrared in nearly-touching metallic nanoparticles. Singular behavior is observed through a low-frequency dipolar mode that migrates towards infinite wavelength as the separation between particles is reduced. This mode loses strength in the limit of touching particles when the interparticle coupling due to charge buildup at the gap overwhelms the dipolar charge oscillations across the dimer. Unphysical modes are obtained for separated particles, associated with a net charge in each particle. However, these unphysical modes evolve into physical modes and have important consequences when the particles touch and the junction defining the initial contact is made in the overlapping dimer at one point. The latter case also exhibits singular behavior as the junction becomes increasingly sharp. Finally, we explain experimental observations[11] and offer a detailed understanding of the mode structure across the transition from separate to overlapping particles in metallic dimers.

The present work focuses on the case where the contact occurs at one physical point as, for example, in dimers made from spherical particles. A different scenario is presented by particles contacting at flat ends as, for instance, two halves of a nanorod cleanly cut at the center. We find that singular redshifts, similar to the shifts found for the spherical-particle dimers discussed in this work, occur as two flat ends touch. However, smooth behavior is obtained after the particles with flat ends merge and overlap (i.e., the modes evolve as expected for a rod of shrinking length). After we present our results, we will return to this point to emphasize the critical influence that the geometry of the contact has on mode evolution to and through contact.

Our results are obtained from a full numerical solution of Maxwell's equations using the boundary element method [20] (BEM), which is based upon discretization of surface charges and currents that are calculated self-consistently to guarantee that the boundary conditions of the electromagnetic field are satisfied. This goes beyond previous non-retarded solutions based upon mode-matching expansions [17,18]. We carry out convergence to the point where BEM calculations for non-touching particles cannot be distinguished on the scale of the figures shown here from calculations obtained by a multipole multiple-scattering formalism [21]. It should be noted that we are using a common calculational scheme to approach the point of touching from both the separated-spheres and the overlapping-spheres sides. Such a study allows us to gain a full understanding of the transition point that has not been previously reported. We focus on dimers made from spherical Au nanoparticles. The dielectric response is described using tabulated empirical data for Au [30]. In the experiment [11], the dimers studied were circular disks, touching on their side along a single line. We consider dimers made of spherical particles here because our BEM simulations exploit cylindrical symmetry to reduce the computational times. We consider the effects of the geometry of the junction by comparing with the case of a split cylindrical nanorod as the flat ends of the split parts contact because this case also exploits cylindrical symmetry.

## 2. Non-touching particles

The evolution of the optical spectrum of a separated particle dimer is presented in Fig. 1(a) through the imaginary part of the dimer effective polarizability. The dimer is formed by identical spherical gold nanoparticles whose surfaces are separated by a distances $d$, ranging from one sphere radius $a$ (lower curve) down to nearly touching particles ($d/a$=0.0005 in the upper curve). Here, we consider spheres with $a$=60 nm. Similar results are obtained for slightly larger particles and for arbitrarily smaller ones. The latter corresponds to the electrostatic limit, in which case extinction spectra are dominated by absorption. For $a$=60 nm spheres however, elastic scattering dominates the extinction spectra, while absorption amounts to ~10-35% of the total cross section (the actual percentage depends strongly on both $d$ and the resonance wavelegnth), showing features that mimic quite closely those of extinction spectra. When the incident electric field is polarized perpendicular to the dimer axis, the spectral shapes are weakly dependent on particle separation (not shown) resembling those of isolated particles, except for a comparatively small blue shift [10-12]. The more interesting situation, considered in Fig. 1 and throughout this work, is for the incident electric field polarized along the dimer axis. For $d$=$a$ the interaction between the particles is weak and the spectrum is dominated by the single-particle dipole mode, which shifts towards the red due to interparticle coupling as $d$ decreases. The shift becomes dramatic near touching, where this mode goes continuously towards infinite wavelength. Actually, the modes in this near electrostatic limit are qualitatively well described by Laplace's equation [31-3], which can be recast as an eigensystem problem that depends exclusively on the geometry and where the eigenvalues $\Lambda$ are related to the actual dielectric properties (e.g., the gold dielectric function $\varepsilon_{Au}$ in our case) as [27,34]

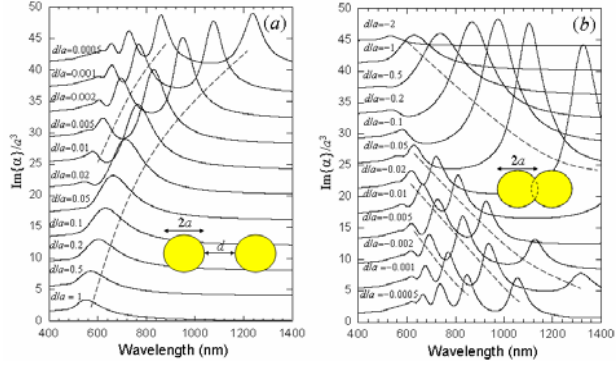

Fig. 1. (a) Wavelength dependence of the imaginary part of the polarizability of the dimer formed by two spherical gold particles ($a$=60 nm) for different distances between their surfaces $d$. The applied electric field is polarized along the interparticle axis. (b) Same as (a) for a gold particle consisting of two overlapping spheres ($d$<0). Dashed curves have been added to guide the eye through the evolution of spectral features with varying $d$.

$$\Lambda = \frac{\varepsilon_{Au}+1}{\varepsilon_{Au}-1}. \qquad (1)$$

We find that one of the eigenvalues evolves as $\Lambda-1 \sim -\sqrt{d}$ as $d$ becomes small compared to the particles radius. Therefore, this eigenvalue approaches $\Lambda=1$ as the particles touch, a value that can only be reached in the $\mathrm{Re}\{\varepsilon_{Au}\} \rightarrow -\infty$ limit, that is, in the long-wavelength limit where the metal exhibits nearly perfect conductor behavior.

This behavior for non-overlapping particles is summarized in the right part of the contour plots of Fig. 2, which show the scattering cross section as a function of wavelength and distance between spherical particles. The electric near-field distribution has been represented in Fig. 3 for selected points of Fig. 2, as indicated by labels in both of these figures. At large separation (I) the electric field near each particle shows a dipole-like pattern, with intense fields in the directions of the interparticle axis. When the distance is decreased (J and K), the electric field is enhanced in the interparticle region in response to huge buildup of induced charge (see the accompanying 3D plots in Fig. 3).

Obviously, oscillations driven by external light in electrically separated particles must respect charge neutrality within each particle. In the noted dipole mode, the charge pileup in each particle near the gap region has a spatial width along the sphere surface of the order of $(ad)^{1/2}$ that is compensated by a smooth distribution of the opposite charge along the rest of the sphere surface. As $d$ decreases and the pileup increases, the magnitude of this smooth charge distribution is enhanced, leading to larger dipole strength of the dimer as a whole. Eventually, at small enough separations the dipole mode disappears because the end-to-end response across the particle pair cannot be driven when the charge coupling across the gap becomes too large. This disappearance is clearly seen in experiment [11]. The onset of this disappearance is already noticeable in the upper curve of Fig. 1(a), and smaller separations (not plotted) show clearly this effect. For separated particles, the net dipole of the dimer is mainly associated to the smooth charge away from the gap, because the distance between opposite charges is then of the order of the dimer length (i.e. several $a$), as compared to the small distance $d$ for the contribution of the charge pileup at the gap. Charge neutrality within each sphere is at the heart of this argument. *Intraparticle* charge neutrality is no longer required when the two particles overlap (i.e. each half of the overlapping dimer can have a net charge). This has important consequences for the mode evolution when contact is made.

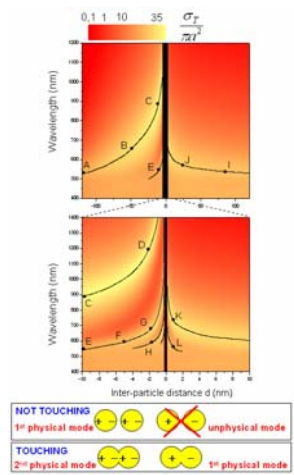

Fig. 2. Contour plot showing the scattering cross section for two spherical gold particles as a function of the separation between their surfaces d (horizontal axis) and the light wavelength (vertical axis). Negative values of d correspond to overlapping spheres. The incoming electric field is polarized parallel to the line connecting the particle centers. The particle radius is $a$=60 nm. The cross section has been normalized to the projected area of one particle, $\pi a^2$. Solid curves are intended to guide the eye through the cross section maxima. The lower inset illustrates how unphysical modes become physical and dominant after touching (A-B-C curve).

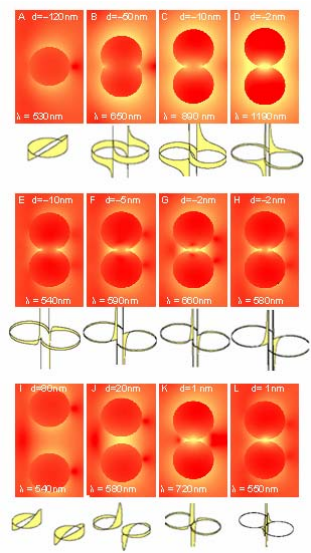

Fig. 3. Near-field maps and the corresponding induced surface charge distributions for two neighboring gold spheres as a function of their separation $d$ and wavelength $\lambda$ corresponding to selected points of Fig. 2, as indicated by labels A-H for overlapping spheres and I-L for separated spherical particles. The sphere radius is $a$=60 nm. The contour plots show the squared electric field in a plane that contains both sphere centers and that is parallel to the incident light direction (the light is coming from the left) and to the incident electric field polarization. The induced surface charge distribution corresponding to the excited modes under consideration is represented in the accompanying 3D plots.

The presence of a low-frequency mode appears to be a property of all metallic systems approaching toward a single point of contact, where a Λ→1 eigenvalue is encountered in the long-wavelength limit, associated with charge pileup at the gap region and charge redistribution in the rest of each particle that guarantees intraparticle charge neutrality. The magnitude of the coupling of this mode to external light (e.g., via a net dimer dipole, as in our case) will depend on the details of the remaining part of the geometry, and further research in the design of the latter is needed to optimize this coupling. Similar behavior occurs when there is a flat surface of contact.

As the particles approach, the dominant dipolar feature becomes narrower, presumably as a result of lower absorption of the gold at larger wavelengths [see the evolution of the right-most peak in Fig. 1(a)]. At the same time, higher-frequency modes take up some of the overall light scattering strength in order to preserve the frequency sum rule. These modes, which become apparent at small separations as distinct spectral features that also shift toward the red, are the result of strong interaction between single-particle multipoles. Their associated charge pileup is characterized by strong oscillations near the interparticle gap (see charge distribution in plot L of Fig. 3). Each of these modes first increases in strength as $d$ decreases, and then eventually decreases for even smaller $d$. This occurs sequentially for each successive shorter wavelength mode.

The eigenvalue problem for a non-touching dimer also reveals the presence of eigenmodes that are associated with spheres that have a net charge [27]. Such modes can exist where the whole dimer is neutral. Other such modes can exist where the dimer has a net charge. These modes must be disregarded as unphysical for excitations of a neutral dimer made from separated neutral particles driven by external light [27]. In particular, one of the unphysical modes where the dimer is neutral but each sphere is charged occurs at a wavelength longer than that of the dominant physical dipole mode. The fate of this unphysical neutral dimer mode changes dramatically when the particles overlap ($d<0$) and electrical contact between the particles is established.

## 3. Overlapping particles

When the particles overlap, it is the interaction between charge that piles up within the anti-wedge formed at the junction where the particles merge (rather than within a gap region) that drives the excitation spectra [see Fig. 1(b)] that evolves after contact. As the particles overlap, this anti-wedge widens (i.e., the overlap region widens) and the modes blue shift, in part, due to mode decoupling. For nearly complete overlap, the dimer acts, effectively, as a single elongated particle and the modes of the dimer behave like the modes of a nanorod, blueshifting as the long axis of the elongated particle decreases. One might expect that the charge buildup at the physical gap of separated particles would neutralize as soon as the gap was shorted out. Surprisingly, the charge remains on the edges of the anti-wedge, leading to electric field distributions that indicate a singular behavior after the particles overlap (see plots A-H in Fig. 3). This behavior is accompanied by modes pushed relatively far into the infrared (see the left part of the plots in Fig. 2). For the lowest-frequency mode, there is a net electrical charge in each half of the dimer (see plots A-D in Fig. 3); that is, each particle is charged. This is a physical situation that becomes allowed by the electric contact at the overlap region, as illustrated in the lower inset of Fig. 2: the left part of the inset shows schematically two separate particles, each with dipole-like charge polarization that give rise to a net-dipole mode as the two particles overlap; however, the unphysical mode with each particle charged in the separate-particle dimer on the right hand side of the inset gives rise to a physical mode that corresponds to the infrared peak that emerges after touching.

Importantly, there is an apparent "jump" between the lowest allowed mode before and after touching with the lowest mode further redshifted after touching. This same jump is seen experimentally [11]. This jump suggests that the lowest-frequency mode evolves after touching from the unphysical, charged modes of the separated dimer, because these unphysical modes lie at lower frequency than the first allowed physical dipole mode of the separated dimer.

Therefore, this low-frequency unphysical mode for charged, separated, spherical particles becomes a physical mode after the particles touch at a single point and leads to extremely low-frequency oscillations right after touching. This shows that the dominant low-frequency modes of non-touching and touching dimers are distinctly different. For non-touching spheres, the lowest-frequency *dipolar* mode, which arises from coupling between dipole modes of each sphere, is charge neutral in each sphere. For touching spheres, the lowest-frequency mode is a *true* dipolar mode with a net charge on each particle. At the transition point with just touching particles, the lowest-frequency mode of the separated dimer must transition into higher-frequency modes of the overlapping dimer while the lowest-frequency mode of the overlapping dimer appears suddenly out of the unphysical modes of the separated dimer. The anti-wedge effect eventually disappears as the overlap grows and the response again becomes that expected for a single particle. The transition to the single-particle limit becomes apparent for overlapping particles when the long-wavelength mode grows relatively in strength enough to again be the dominant mode [see Fig. 1(b) for $d/a \leq -0.2$].

In contrast, when particles with flat ends touch, the dipole mode jumps in the opposite direction to higher frequency. This further shows the critical influence of the geometry of the junction. When particles with flat ends touch, the charge at the junction is neutralized, again in contrast to the charge buildup that remains at a point contact. When flat ends touch, the lowest physical dipole mode of the overlapping particle evolves from the spectrum of physical modes of the separated dimer.

The overlapping dimer of spherical particles exhibits a rich structure of higher-frequency modes for small overlaps [see curves for $-0.1 \leq d/a \leq -0.0005$ in Fig. 1(b)], also shifting towards the blue as the overlap increases. These modes show strong electric field enhancement at the anti-wedge region, which is accompanied by charge pileup with more nodes as compared to the lowest-frequency mode (see plots E-H in Fig. 3). Actually, the induced charge is anti-symmetric with respect to the plane that separates both particles for the illumination conditions under consideration, and this is what makes a net dimer dipole. The charge in one of the halves of the dimer does not change sign in the lowest-frequency mode (A-D in Fig. 3), whereas the charge has one node in each particle in the second-lowest-frequency mode (E-G), two intraparticle nodes in the third-lowest-frequency mode (H), and so on.

Further insight into the nature of these modes can be gained by making a connection between touching dimers and dumbbells. Fig. 4 presents the wavelength of absorption features of dumbbell particles as a function of their length $L$. In the touching dimer limit ($L \rightarrow 0$), two dominant absorption peaks are observed, one of them corresponding to the right-most dashed curve of Fig. 1(b) (actually, with 1-nm smoothing radius at the junction; see Fig. 5 for a discussion of the role of smoothing). As the dumbbell length increases, this mode shifts towards the red, in what appears to be a standing-wave-like behavior associated to an induced charge density pattern as shown in the central inset of Fig. 5. For large $L$'s, this mode broadens as a consequence of a decrease in its radiative lifetime, which is accompanied by a significant increase in extinction cross section (not shown). The other feature occurs near 500 nm and is relatively independent of $L$. In fact, it is very close to the dipole Mie resonance of an isolated gold sphere. Further inspection into the associated charge density of this lower-wavelength mode reveals the excitation of dipole-like modes localized at each of the spherical ends of the dumbbell for large $L$'s. As $L \rightarrow 0$, strong interaction between spherical ends produces broadening of this mode, which mixes with other multipolar components to yield the complex spectra shown in Fig. 1(b). From here, one may conclude that the longest-wavelength feature of the touching dimers of Fig. 2 results from the evolution of a standing-wave mode of the dumbbell, whereas the shorter-wavelength peaks separate neatly into Mie resonances similar to those of isolated spheres as the dumbbell length increases.

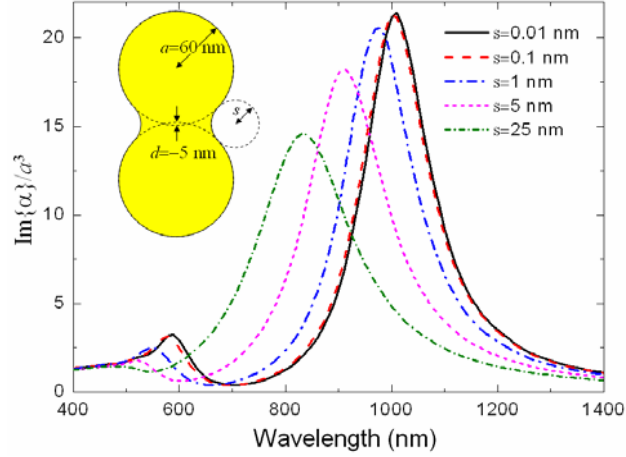

Fig. 4. Resonant wavelengths of standing and localized modes in gold dumbbell particles as a function of the length $L$ of the central rod for two different values of the overlap distance $d$ at $L$=0. All geometrical parameters are given in the insets.

**4. Sharp versus smooth junctions**

In general, singular behavior is connected to sharp edges and junctions, as for a metallic wedge where the plasmon modes are discrete for a rounded edge but a continuum for a sharp wedge [35]. Anomalous response in metallic particle arrays has also been long observed at the point of percolation, where the dielectric constant rises to infinite positive values right before touching [36-38] and fluctuations of the electrical conductivity exhibit a divergent behavior [39]. In our case, touching particles and particles overlapping with sharp anti-wedge structure provide, as well, singular situations that are accompanied by arbitrarily-low-frequency modes.

We have explored the transition from smooth to sharp junctions by rounding the anti-wedge junction of overlapping particles with a small radius of curvature $s$, as shown in the inset of Fig. 5. For any fixed value of $d$<0, the scattering spectra converge smoothly to the results presented in Figs. 1−3 in the $s\rightarrow0$ limit. An example of this convergence is shown in Fig. 5 for $a$=60 nm and $d$=−5 nm. The main peak observed in this figure lies on the CD curve of Fig. 3.

**5. Further discussion and conclusions**

Our theory provides insight into the nearly touching regime by virtue of numerical simulation, resulting in the detailed description of the evolution of the modes and induced charge (see Fig. 3). Our results explain the main features of the behavior observed in Ref. [11] in the optical response of lithographically patterned gold dimers. Namely: (i) the non-touching particles exhibit a dominant feature at large separations that is related to dipolar polarization of each particle and that shifts towards the infrared as the particles get closer, the response of this mode first increases as $d$ is reduced and then decreases for even smaller $d$; (ii) for overlapping particles, a low-frequency mode of dipolar nature shows up that shifts towards the infrared when the overlap region is reduced and this shift is more pronounced than in separated particles with a jump further to the red at contact; (iii) higher-frequency modes of higher multipolar character appear on both sides of the touching limit; (iv) small changes in the region close to interparticle contact in the nearly touching and barely overlapping limits produce huge variations in the optical spectra of the dimer; and (v) when the light polarization is perpendicular to the dimer axis, our calculations (not shown) indicate that these variations are minor, with a main feature showing up near the wavelength corresponding to an isolated particle.

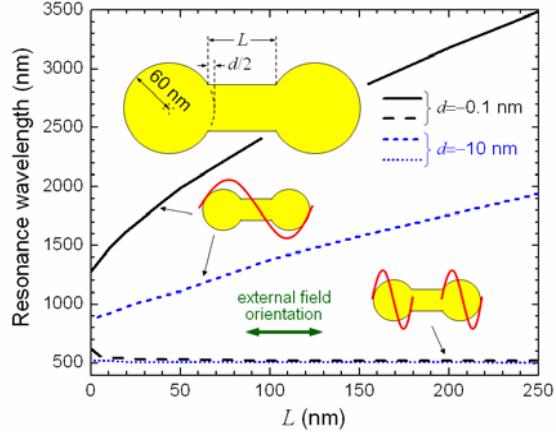

Fig. 5. Wavelength dependence of the imaginary part of the polarizability of an overlapping dimer with a smooth junction for sphere radii $a$=60 nm and overlap parameter $d$=−5 nm. The junction is smoothed by a toroidal surface of inner radius $s$ as shown in the inset. Various values of $s$ have been considered.

We have assumed so far that the optical response is local and well described by frequency-dependent dielectric functions. However, non-local effects come about when the size of the particles is comparable to the inelastic mean free path of the valence electrons that participate in the plasmon resonances [2]. Although non-local effects are negligible for spherical metallic particles of several tens of nanometers [40], like the ones considered here, the singular character of the point of contact between particles can be severely modified by those effects [40], serving to cut off the singularity. Moreover, charge leakage between two very closely spaced but separated particles further complicates the non-local response. More theoretical insight into finite-size effects near the contact point is still needed [26]. In addition to this, the near touching limit should be very sensitive to a realistic density profile at the particle surface, which will presumably impose a lower bound to the actual overlap distance $d$ of the just-touching particles of the order of the spill out distance of the static valence electron density (~0.2 nm).

The smallest gaps that we have considered are unphysically narrow (less than a lattice spacing). However, our discussion for the shortest gap and overlap distances takes physical relevance beyond academic discussion when considering metallic behavior ($\varepsilon \ll -1$) at optical phonon frequencies in alkali-halides, where we have observed similar red shifts of the modes in overlapping and non-overlapping 4-μm KCl spherical particles near the ~50-μm phonon mode of KCl spheres. Fig. 6 illustrates similar behavior as for smaller gold particles, and in particular multipolar peaks are observed for a ratio $d/a$=0.017, which corresponds to a nearest-neighbor distance of $d$=33 nm, enough to guarantee the adequacy of the local response assumption). Alkali-halide particle dimers of several microns in diameter should offer excellent systems with which to probe experimentally the ideas covered in this work in the far infra-red domain.

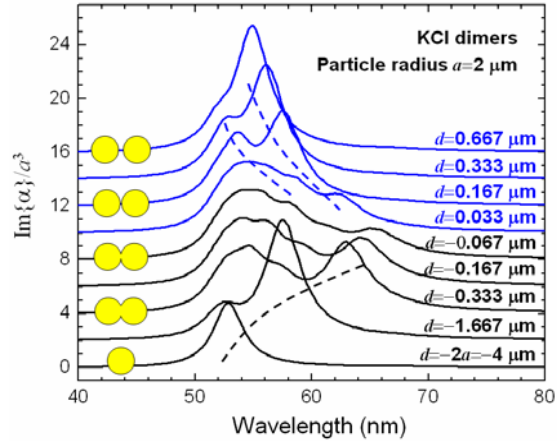

Fig. 6. Wavelength dependence of the imaginary part of the polarizability of the dimer formed by two spherical KCl particles of radius $a=2$ μm for different distances between their surfaces $d$. The applied electric field is polarized along the interparticle axis. $d<0$ stands for overlapping particles. Dashed curves have been added to guide the eye through the evolution of spectral features with varying $d$.

The enhancement of the electromagnetic field in the gap region of adjacent particle pairs has been used to perform surface-enhanced Raman spectroscopy (SERS) [13,23], where the magnitude of the light scattering signal scales with the fourth power of the electric field. Our work shows that particles overlapping over a small region can produce more dramatic effects, therefore suggesting new ways of improving SERS sensitivity at anti-wedge sites. This can have direct application to the detection of single molecules (e.g., DNA, proteins, or bio-molecules) that can attach to markers (e.g., bio-markers) initially deposited at anti-wedge sites.

In summary, we have analyzed the nearly touching regime of metallic nanoparticle dimers and explained how long-wavelength resonances occur both for approaching particles and after particle overlap, separated by singular behavior right when the particles start touching. Both of these regimes are characterized by large pileup of induced charge at the gap or overlap region. However, the charge at the gap and the lowest-frequency modes are distinctly different for these two regimes and are not connected via the transition through the limit where the spheres are just touching. Our results help explain recent experiments performed on nanolithographically prepared particles [11] and provide relevant information on the structure of plasmon modes at the contact point between metallic particles that could be used to design new plasmon guides based upon overlapping particles, as an extension of wave guiding in non-overlapping dimers [41-43]. Furthermore, dimers with small overlap regions produce larger enhancement than non-touching particles in close vicinity, thus suggesting a strategy for increasing the degree of enhancement in applications such as SERS.

**Acknowledgements**

FJGA wants to thank Carlos Pecharromán for helpful and enjoyable discussions. This work was supported in part by the University of the Spanish Ministerio de Educación y Ciencia (FIS2004-06490-C03-02 and NAN2004-08843-C05-05).